\documentclass[longauth]{aa}  

\usepackage{graphicx}
\usepackage{txfonts}
\usepackage{threeparttable, tablefootnote}
\usepackage{hyperref}
\hypersetup{colorlinks = true,linkcolor= blue,urlcolor=blue}
\usepackage{chemformula}
\usepackage{placeins}
\usepackage{float}
\usepackage{stfloats}
\usepackage{soul}
\hypersetup{%
citecolor=blue
}
\long\def\ce#1{\ensuremath{#1}}

\begin{document} 

\title{Mapping of interstellar ices with JWST in protostellar envelopes:}
\subtitle{Spatial variations of $^{13}$CO$_2$, $^{12}$CO$_2$, CO, OCN$^-$ and H$_2$O ice}

\author{N. G. C. Brunken\inst{1}, E. F. van Dishoeck\inst{1,2}, A. C. A.  Boogert\inst{3}, P. Nazari\inst{4}, Y. Chen\inst{2}, J. Santos\inst{5}, H. Beuther\inst{6}, A. Caratti o Garatti\inst{7}, T. Megeath\inst{8} K. Slavicinska\inst{1,9}, C. Gieser\inst{6}  H. Tyagi\inst{10}, L. Tychoniec\inst{1}, M. McClure\inst{1}, V. J. M. le Gouellec\inst{11},Y. Yang\inst{12}, L. W. Looney\inst{13,14}, N. J. Evans \inst{15} and M. Narang\inst{10}}

\institute{Leiden Observatory, Leiden University, 2300 RA Leiden, The Netherlands \\
\email{brunken@strw.leidenuniv.nl} %1
\and 
Max-Planck-Institut f\"{u}r Extraterrestrische Physik, Gie{\ss}enbachstra{\ss}e 1, 85748 Garching, Germany %2
\and
Institute for Astronomy, University of Hawai’i at Manoa, 2680 Woodlawn Drive, Honolulu, HI 96822, USA %3
\and
Institute of Astronomy, University of Cambridge, Madingley Rd, CB3 0HA, Cambridge, UK %4
\and
Center for Astrophysics, Harvard \& Smithsonian
60 Garden St, MS 16
Cambridge, MA 02138, USA/51 Pegasi b Fellow. %5
\and
Max Planck Institute for Astronomy, Heidelberg, Baden Wuerttemberg, Germany %6
\and
INAF-Osservatorio Astronomico di Capodimonte, Italy %7
\and
Department of Physics and Astronomy, The University of Toledo, 2801 West Bancroft Street, Toledo, OH 43606, USA %8
\and
Laboratory for Astrophysics, Leiden Observatory, Leiden University, PO Box 9513, 2300 RA Leiden, The Netherlands %9
\and
Tata Institute of Fundamental Research, Mumbai 400005, India %10
\and
NASA Ames Research Center, Space Science and Astrobiology Division M.S 245-6 Moffett Field, CA 94035, USA %11
\and
RIKEN Cluster for Pioneering Research, Wako-shi, Saitama, 351-0106, Japan %12
\and
Department of Astronomy, University of Illinois, 1002 W. Green St., Urbana, IL, 61801 %13
\and
National Radio Astronomy Observatory, 520 Edgemont Rd., Charlottesville, VA 22903, USA %14
\and
Department of Astronomy, The University of Texas at Austin, 2515 Speedway, Stop C1400, Austin, Texas 78712-1205, USA %15
}

\titlerunning{}
\authorrunning{N.G.C. Brunken et al.}

\date{Received; Accepted}

\abstract
{The composition of planets and their atmospheres is largely determined by the environment in which the planet accretes material. This environment is pre-determined during early stages when the protostar is still forming and impacting the complexity of the material in the envelope. Charting the structure of protostellar envelopes is therefore crucial for understanding how the chemistry is linked to the physical processes occurring in stellar nurseries, and JWST offers a unique opportunity to perform spatial mapping at sub-arcsec resolution.  }
{We aim to explore spatial variations of the main ice constituents in the envelope of the low-mass protostar L1527 and the high-mass protostar IRAS 20126+4104 using JWST observations. }
{We use JWST NIRSpec IFU observations to extract spectra at each pixel in the IFU of $^{13}$CO$_2$,$^{12}$CO$_2$, CO, OCN$^-$ and H$_2$O ice features. We create optical depth and spatial abundance maps to study the distribution of the ices and use the $^{13}$CO$_2$ ice feature, a known thermal tracer, to identify regions of high temperatures where the ices are thermally processed.}
{Our findings show that the distribution of ices in both L1527 and IRAS 20126+4104 is not uniform, with both envelopes showing localized temperature zones and substructures. In L1527,  CO$_2$ ice segregation and OCN$^-$ enhancement with respect to CO$_2$ are observed in the warm central disk lane close to the protostar. In IRAS 20126+4104, the warm temperature zones are observed mostly towards the shocked region in the NW outflow where dense material is interacting with the precessing jet as well as close to the central source. The CO$_2$ ice is segregated at these locations and we observe sublimation of CO ice, as well as strong gaseous CO rotation vibrational lines, especially in the shocked NW outflow. Close to the central protostar we see an additional enhancement of OCN$^-$ with respect to CO$_2$ that also points towards elevated temperatures and crystalline H$_2$O ice is also detected at a few locations that coincide with shock knots in the outflow.} 
{In both low mass and high mass protostar the findings indicate that the distribution of ices and their physical structure are linked to the temperature structure of the envelopes. The envelope of IRAS 20126 shows a more complex structure, where interacting material results in localized regions of high temperatures and denser regions that harbor cooler ices. L1527 in contrast appears to be a more structured system with heated warm ices observed mostly towards the central disk regions.}

% there is a specific list to choose from for A&A

\maketitle

%-------------------------------------------------------------------

\section{Introduction}

The creation of worlds involves an intricate sequence of events, each governed by numerous processes that shape the chemical makeup and structure of the planetary systems \citep{mumma2011chemical,oberg2021}. During the prestellar and protostellar stage, the initial ingredients of planetary matter are established with the formation of interstellar ices, the chemical seeds of more complex chemistry \citep{herbst2009, caselli2012,boogert2015observations}. Physicochemical processes triggered by the accreting system can subsequently transform this initial composition. 

Indeed, radiation and heating from the infant star as well as shocks from jets and outflows can create temperature gradients and localized warm regions in the envelope and disk. This in turn can impact the distribution and chemical complexity of the material \citep{evans1999,schoier2002, dunham2014,murillo2018,tobin2024}. The young system as a whole can also be affected by its local environment. Massive stars, for instance, are known to form in bustling neighborhoods that are subjected to strong UV-fields and high-energy particles that easily penetrate clouds and circumstellar medium, affecting the chemistry of the envelope and disk. In addition, these regions have a high star population density, which often leads to young stellar companions that interact dynamically \citep{zinnecker2007,sana2012, duchene2013, motte2018}, thus having additional consequences on both the physical structure and chemistry of the individual systems. In contrast, the conditions in low-mass stellar nurseries are generally less violent than their high-mass counterparts \citep{beuther2025}, which could lead to differences in the chemical structure of their envelopes. This raises two questions: to what extent is the chemistry affected by these protostellar processes and are there observable differences in the distribution of matter inside the envelopes of these systems?

Previous observational facilities have enabled detailed characterization of the molecular inventory in these young systems in both gas- and ice-phase \citep{jorgensen2020,hoff2026,ceccarelli2023,rocha2024,chen2024}. Gas-phase observations and modeling studies in particular have shed light on the morphology and kinematics of protostellar systems that directly affect circumstellar matter. In addition, these studies were successful in extracting temperature profiles and establishing the location of snowlines in various protostellar envelopes \citep{jorgensen2002,jorgensen2011, hoff2018b,Hsieh2019}.  In the solid state, infrared observatories have shown that protostellar systems are teeming with ices that vary in relative abundance depending on the line of sight \citep{gibb2004,oberg2011spitzer,boogert2015observations}. In addition, their structure and composition can be altered by thermal processes such as segregation and sublimation of the highly volatile species  \citep{pontoppidan2008c2d,oberg2011spitzer,brunken2025}.

Although these studies collectively provide a multifaceted view of the overall architecture of protostellar systems and the available material for planet assembly, the link between the envelope structure and the spatial distribution of the numerous chemical species remains largely unexplored. This is especially true for interstellar ices, where spatial mapping in clouds and protostellar envelopes has only recently become feasible with the James Webb Space Telescope (JWST) at high (subarcsec) angular resolution and spectral \citep{tyagi2025,Petrashkevich2026}. Before JWST, ice mapping studies were mainly focused on infrared sources behind dark clouds and envelopes and were limited to a number of species such as CO and CH$_3$OH that could be observed from the ground \citep{pontopiddan2004,pontoppidan2006,sonnentrucker2008}. 

The new technology provided by the JWST NIRSpec Integral Field Unit (IFU) now offers an opportunity to investigate the location and physical structure of ices in the envelopes around protostars spanning a range of masses and luminosities \citep{yang2022,brunken2024a,slavicinska2024,nazari2024}. The spatial distribution and ice abundances of H$_2$O, CO and CO$_2$, the main ice constituents, as well as ionic species such as OCN$^-$ can be determined for the first time in young systems and be directly connected to the physical structure of the envelope. Moreover, the high spectral resolution allows us to study variations in ice band profiles and hunt for thermal indicators \citep{brunken2024a,brunken2025}, enabling the identification  of temperature zones in protostellar envelopes from a new perspective. Segregation of CO$_2$ ice, sublimation of CO ice, and enhancement of OCN$^-$ ice with respect to other species can all point to warmer environments \citep{boogert1999iso,pontoppidan2008c2d,brunken2025,nazari2026}.
By combining these new ice studies with previous findings from gas-phase observations, we can obtain a comprehensive view of the structures and chemical abundance in protostellar envelopes, which are important for our understanding of planet formation since the planet composition is determined by the material it accretes and, therefore, its formation location. 

In this work, we present JWST NIRSpec IFU spatial maps of the low-mass protostar L1527 (L = 3 L$_{\odot}$, M = 0.45 M$_{\odot}$) and the high-mass protostar IRAS 20126+4104 (hereafter IRAS 20126) (L = 10$^4$ L$_{\odot}$, M = 12 M$_{\odot}$ ) as part of the JOYS+ \citep{beuther2023, dishoeck2025} and Investigating Protostellar Accretion (IPA) collaborations \citep{federman2023,narang2024,rubinstein2023} (see also Appendix \ref{Appendix:A} for additional source properties). To the best of our knowledge, this is the first ice mapping study of a high-mass protostar. The observations include the H$_2$O stretching mode (3 $\mu$m), the $^{12}$CO$_2$ \ce{\nu_3} stretching mode (4.27 \ce{\mu}m), the $^{13}$CO$_2$ \ce{\nu_3} stretching mode (4.39 \ce{\mu}m), the OCN$^-$ stretching mode (4.60 $\mu$m) and the $^{12}$CO stretching mode (4.67 \ce{\mu}m) (Figure \ref{fig:L1527_int} right Panel). The analysis is focused on NIRSpec observations because they provide extended scattered light emission against which the ice absorption can be observed, unlike the dust thermal emission that is obtained with MIRI which is more compact. The paper is structured as follows. In Section \ref{sec:2} we present the observations and methods. In Section \ref{sec:3} we provide the analysis of the maps and the results are discussed in Section \ref{sec:4}. Finally, in Section \ref{sec:5} we provide a summary and the concluding remarks.

\section{Observations and methods}
\label{sec:2}

\subsection{Observations}

The NIRSpec and PRISM observations of L1527 were taken as part of the JOYS+ Cycle 1 GO program (PI: E.F. van Dishoeck, ID: 1960) and the data consist of NIRSpec (2.87 - 5 $\mu$m) and PRISM spectra observed using the G395H modes ($R$ = $\lambda$ /$\Delta$ $\lambda$ = 2700). The IRAS 20126 data were taken as part of the IPA Cycle 1 GO program (PI: T.Megeath, ID: 1802) using the NIRSpec G395M mode ($R$ = $\lambda$ /$\Delta$ $\lambda$ = 1000). Both settings use the integral field unit. Further observational details, data reduction methods, information on the spectral extractions, source properties and coordinates are provided in Section \ref{sec:22}, Table \ref{Tab:1}, and in references \citep{brunken2024a,brunken2024b,gelder2024a,federman2023,rubinstein2023,narang2024,dishoeck2025,devaraj2026}. 

\subsection{Spectral extractions}
\label{sec:22}
To obtain spectral information, we utilize a python script that loops over each pixel in the IFU and extracts a spectrum between 2.87 - 5.27 $\mu$m. Each pixel is 0.1 $\times$ 0.1 arcsecond and we use this spatial information to convert the maps to astronomical unit scales assuming distances of 142 and 1550 pc for L1517 and IRAS 20126, respectively \citep{tobin2012,cesaroni2023}. For both L1527 and IRAS 20126 the pixels have good S/N and undersampling does not appear to be an issue. In fact, the ice features extracted on a pixel-by-pixel basis look similar to the ice features obtained from 3$\lambda$/D cone aperture extractions centered on the infrared source \citep{brunken2024a, brunken2024b, brunken2025}. The signal does decrease towards the outermost pixels and we therefore refrain from doing any analysis with pixels towards the edge of the IFU field of view. 

\subsection{Continuum subtraction}

To convert the spectra on optical depth scale we fit a second order polynomial over the ice features and convert the spectra using $\tau^{obs}_{\lambda}$ = -ln ($F^{obs}_{\lambda}$/ {$F^{cont}_{\lambda}$), where $F^{obs}_{\lambda}$ is the observed flux and $F^{cont}_{\lambda}$ is the flux of the continuum. Local continua are fitted over the $^{13}$CO$_2$ ice band between 4.35 and 4.47 $\mu$m and the OCN$^-$ and CO ice features between the 4.48 $\mu$m and 5.14 $\mu$m. Examples of these fitted regions and details of the uncertainty on these fittings are shown and discussed in \citet{brunken2024a,brunken2024b}. The full spectra are shown on flux scale in the upper right corner of Figure \ref{fig:L1527_int}. The continua are fitted at each pixel and divided automatically. 

While the 3 $\mu$m H$_2$O band and the 4.27 $\mu$m stretching mode of $^{12}$CO$_2$ are also covered in this spectral window and discussed in this work, obtaining continuum-subtracted spectra is not possible. The H$_2$O band is only observed using the G395 which does not cover wavelengths below 2.9 $\mu$m. Consequently, the short-wavelength wing of the band is not available for continuum fitting. For L1527 this can be remedied by using the PRISM data to guide the continuum but since the water feature saturates into the noise level it is not be possible to do an optical depth scale analysis. PRISM data are not available for IRAS 20126. Continuum fitting over the $^{12}$CO$_2$ stretching mode is challenging because this band is sensitive to grain shape and size effects that warp its wing profile \citep{dartois2006spectroscopic,dartois2022influence,dartois2024}. As a result, spatial variations of the 3 $\mu$m H$_2$O band and the 4.27 $^{12}$CO$_2$ ice feature will be presented on flux scale and quantitative analyses on the column densities are not provided.

\subsection{Spatial maps}

Peak optical depth maps are created for $^{13}$CO$_2$, OCN$^-$ and CO by determining the optical depths for each pixel at the peak wavelengths 4.38 $\mu$m, 4.60 $\mu$m, and 4.67 $\mu$m, respectively. The column densities maps are subsequently calculated by integrating under the ice features and using $N$ = $\int\tau_{\nu}$ $d\nu$/$A$, where $\int$$\tau_\nu$ d$\nu$ is the integrated optical depths under the absorption feature, and $A$ is the corresponding band strength of the vibrational mode.

\begin{figure*}[h!]
    % \vspace{0.5cm}
    \centering
     \includegraphics[width=1\hsize]{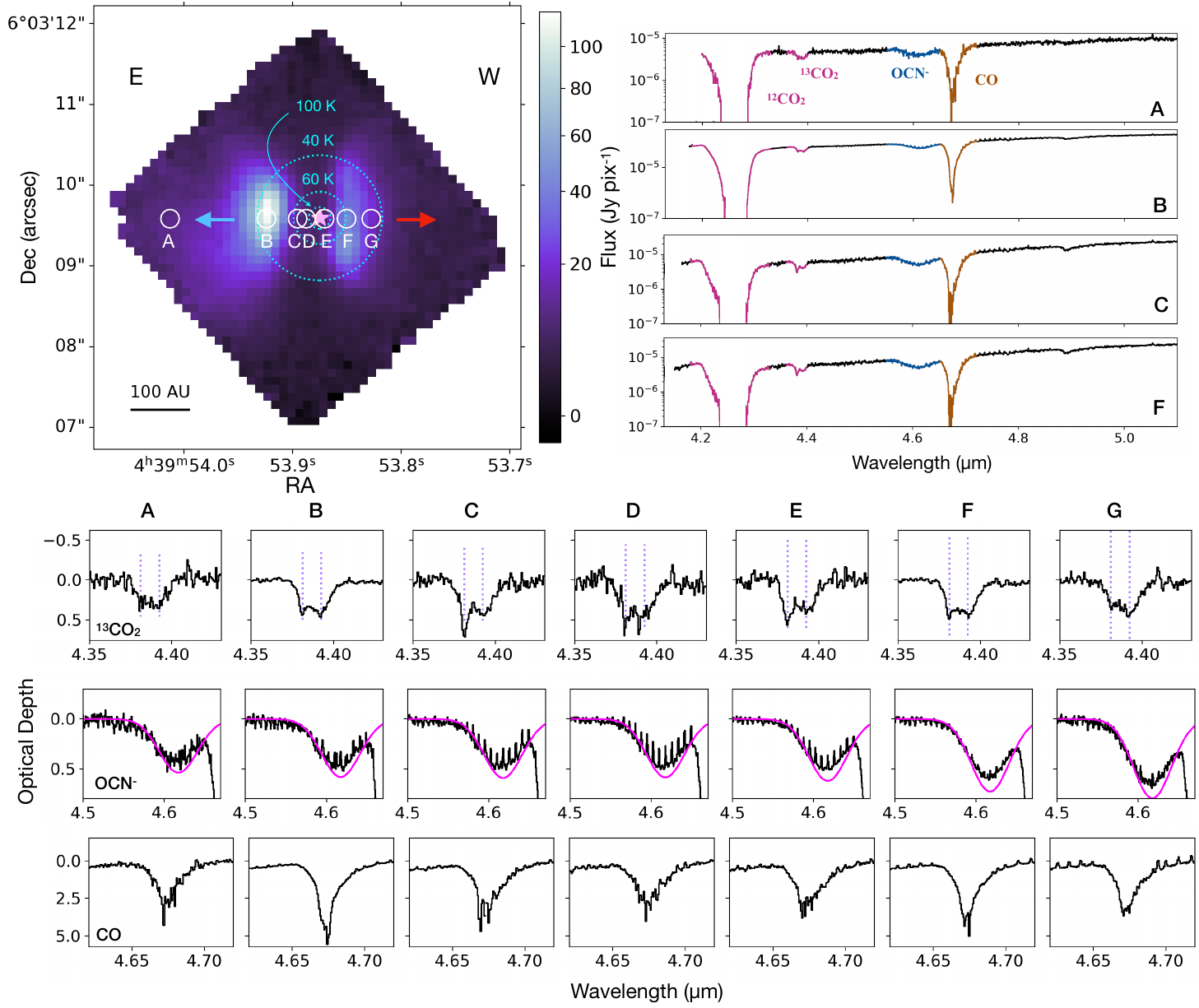}
    \caption{Intensity map of L1527 at 3.8 $\mu$m. The source location is marked with a pink star. The white circles are centered on the pixel where spectra were extracted with the corresponding spectra of $^{13}$CO$_2$, CO and OCN$^{-}$ shown in the panels below. Note different optical depth scales of the ice species. The 4.2 - 5 $\mu$m spectra at the four positions A, B,C and F are shown in the right panel with the ice features denoted in colors (the H$_2$O ice feature is shown due to saturation). The 40 K, 60 K and 100 K dotted circles mark the temperature borders derived from a simple 1D spherical radiative transfer model. The dotted purple lines in the first row show the locations of the 4.38 $\mu$m and 4.39 $\mu$m of peaks of the $^{13}$CO$_2$ ice feature. The pink solid lines show the Gaussian curves fitted over the OCN$^-$ ice feature. The red and blue arrows point the direction of the outflows. }
    \label{fig:L1527_int}
\end{figure*}

For $^{13}$CO$_2$, CO, and OCN$^-$ we use the band strengths 9.98 $\times$ 10$^{-17}$, 1.1 $\times$ 10$^{-17}$ and 1.5 $\times$ 10$^{-16}$ cm molecule$^{-1}$, respectively, taken from \citet{hudson2025,gerakines2023,gerakines2025}. These band newly measured band strengths differ from those used in \citet{brunken2024b} who calculated column densities with band strengths taken from \cite{gerakines1995} and \citet{broekhuizen2005}. The new band strengths differ from the previously used one by a factor 1.15, 1.27 and 0.87 for $^{13}$CO$_2$, $^{12}$CO, and OCN$^-$, respectively. For details on the error budget we refer the reader to \citet{brunken2024b} and \citet{brunken2025}. Since the 4 $\mu$m spectral region is polluted by gaseous CO lines, we opted to fit a gaussian curve over the OCN$^-$ ice band to obtain the column densities. We allow the FWHM to vary between 23 to 26 cm$^{-1}$ and set the boundary for the peak position between 2165.60 and 2165.64 cm$^{-1}$ before fitting the ice feature at each pixel using a simple optimization routine. From the column density maps, we produce CO$_2$/CO, CO$_2$/OCN$^-$ and OCN$^-$/CO ratio maps. Finally, we study the spatial variation of the short-wavelength pure $^{13}$CO$_2$ peak at 4.38 $\mu$m with respect to the long-wavelength peak at 4.39 $\mu$m of $^{13}$CO$_2$ by taking the maximum optical depth at the two fixed positions and computing the ratio. The maps serve as an indication of the segregation of CO$_2$ ice \citep{boogert1999iso,brunken2024a,brunken2025}.

\begin{figure*}[h!]
    % \vspace{0.5cm}
    \centering
     \includegraphics[width=0.7\hsize]{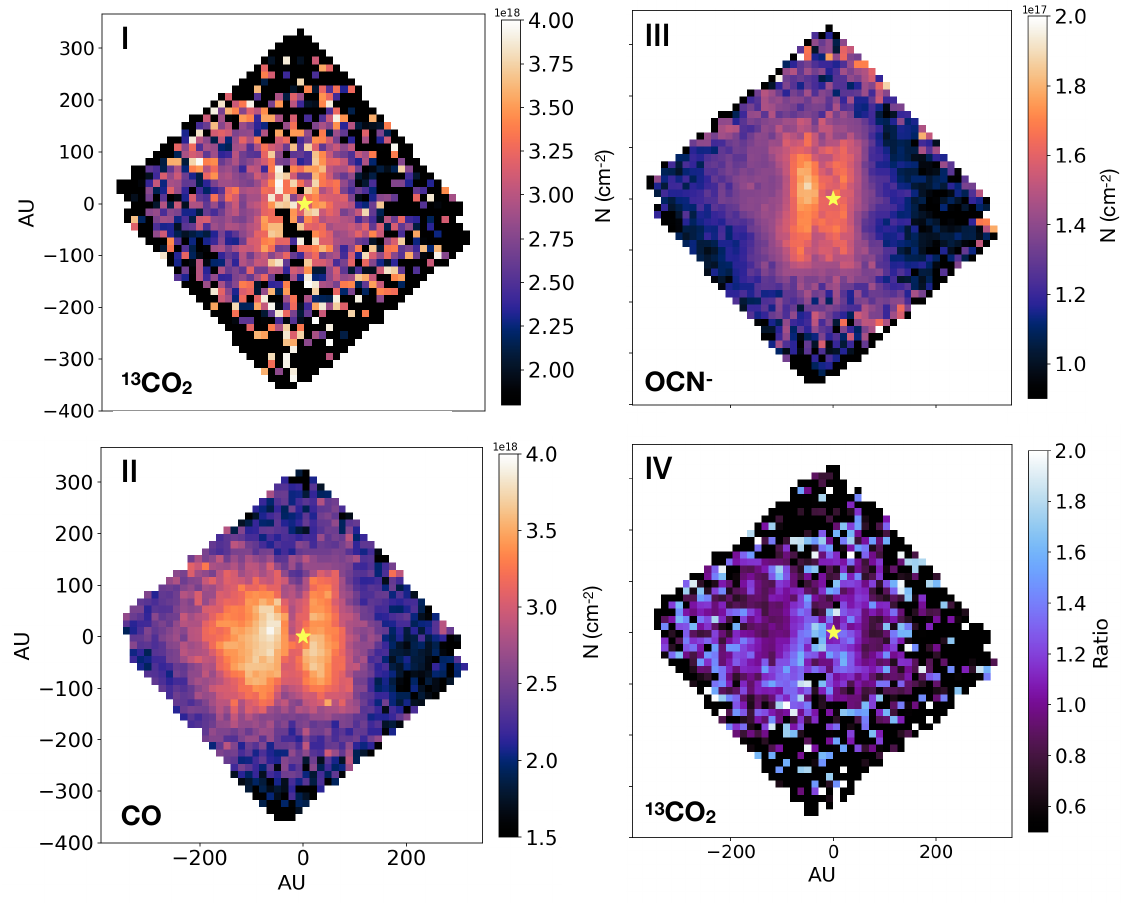}
    \caption{Column density maps of $^{13}$CO$_2$, CO and OCN$^{-}$ ice (panel I, II and III, respectively). $^{12}$CO$_2$ column densities are derived from $^{13}$CO$_2$ column densities. Panel IV shows the ratio between the 4.38 $\mu$m and 4.39 $\mu$m peak of the $^{13}$CO$_2$ ice feature. A higher ratio indicates higher degree of CO$_2$ ice segregation. The axes are shown on astronomical unit scales. Pixelated areas especially in the north and south of the maps as well as the edges are caused by low signal-to-noise levels.  }
    \label{fig:L1527_CD}
\end{figure*}

\section{Analysis}
\label{sec:3}

\subsection{L1527}

The 3.8 $\mu$m continuum map of L1527 is presented in Figure \ref{fig:L1527_int}. The dark lane shows the near edge-on disk in which the central protostar is enshrouded. Emission from the central source is seen in scattered light illuminating the outflow cavities that stretch in the east and west directions \citep{tobin2008,hoff2023}, seen also in mid-infrared atomic and molecular lines \citep{drechsler2026,devaraj2026}. Example spectra at various positions in the envelope are shown on optical depth scale from pixel positions A to G for $^{13}$CO$_2$, OCN$^{-}$ and CO ice. We ran a simple 1D spherical envelope radiative transfer model using RADMC-3D \citep{dullemond2012} to obtain the 40 K, 60 K, and 100 K lines. The parameters used in the models are provided in Table \ref{Tab:1}. We note that although both L1527 and IRAS 20126 have disks
\citep{tobin2012,cesaroni2025}, this is not included in the model. As shown by \citet{sturm2023}, disks can cause scattering effects that affect the path of the photon, but given the large mass of the envelope during these early stages, the photon travels a longer path through the envelope from the point of last scattering, and thus most of the ice absorption is coming from the envelope rather than the disk. The exclusion of disks and outflow cavities in our models also results in spherical contours that are not entirely representative of the envelope and disk scenario. For the purpose of this study however, they serve as an approximation of the various temperature zones and are sufficient to support the observations.

\subsubsection{CO$_2$}

Figure \ref{fig:L1527_CD} panel I shows the column density map of $^{12}$CO$_2$ ice converted from $^{13}$CO$_2$ ice using the standard ISM $^{12}$C/$^{13}$C ratio of $\sim$ 68 \citep{boogert1999iso}. The peak optical depth map of $^{13}$CO$_2$ is presented in Figure \ref{fig:IRAS20126_OD}. Both the column density and the optical depth maps suggest an enhancement of CO$_2$ ices in the central disk region with values reaching $N$ $\sim$ 5 $\times$ 10$^{18}$ cm$^{-2}$ ($^{12}$CO$_2$) and $\tau$ $\sim$ 0.6 ($^{13}$CO$_2$). Lower values are measured towards the outflow cavities with column densities averaging around $\sim$ 3 $\times$ 10$^{18}$ cm$^{-2}$, and the optical depth of the $^{13}$CO$_2$ feature decreasing to $\tau$ $\sim$ 0.3, lower by almost a factor of two. The column densities on average agree with the value reported in \citet{brunken2024b} ( $\sim$ 3 $\times$ 10$^{18}$ cm$^{-2}$ ).

Additionally, we observe a growth of the short wavelength $^{13}$CO$_2$ peak (4.38 $\mu$m), associated with pure segregated CO$_2$ ice, relative to the long wavelength peak  (4.39 $\mu$m), associated with polar CO$_2$:H$_2$O and CO$_2$:CH$_3$OH ices (Figure \ref{fig:L1527_CD} panel IV). The gradual growth of this additional peak is clearly detected between positions B and C, moving in the direction of the disk and the central source (Figure \ref{fig:L1527_int}). Similarly, starting from the west regions, the short wavelength peak appears between positions F and E, growing stronger towards the central source. Figure \ref{fig:L1527_CD} panel IV shows that the pure CO$_2$ feature is mostly observable in this central disk region and along the cavity walls.

\begin{figure*}[h!]
    % \vspace{0.5cm}
    \centering
     \includegraphics[width=1\hsize]{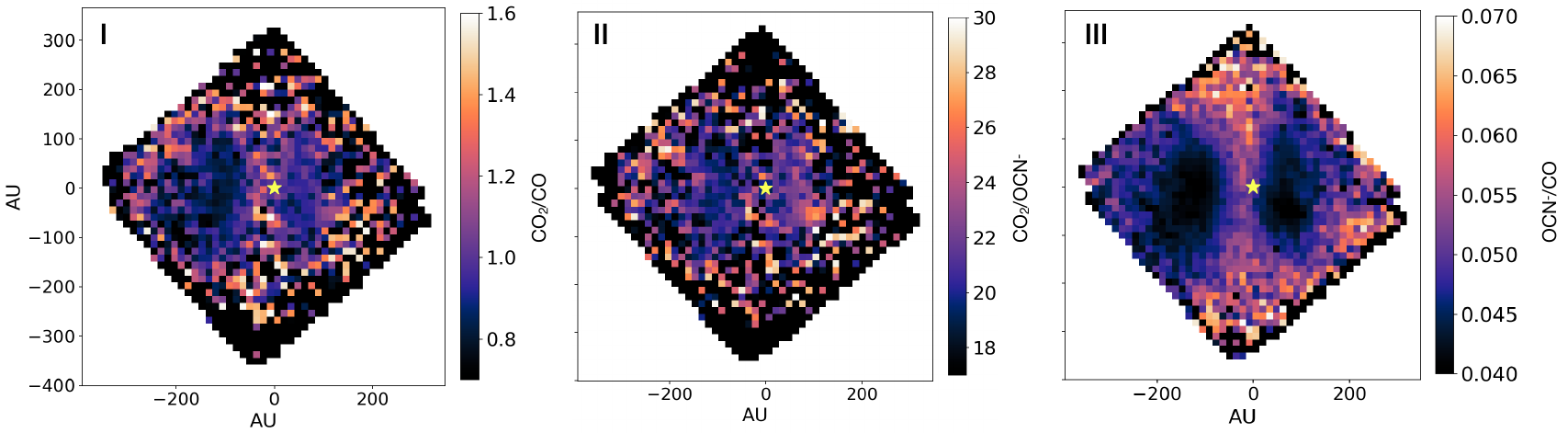}
    \caption{Column density ratio maps for CO$_2$, OCN$^-$ and CO ice. The $^{12}$CO$_2$ column densities were determined using the standard $^{12}$C/$^{13}$C ratio of $\sim$ 68.  }
    \label{fig:L1527_abun}
\end{figure*}

\subsubsection{CO}

The column density and optical depth maps of CO ice are shown in panels II of Figures \ref{fig:L1527_CD} and \ref{fig:L1527_OD}, respectively. In contrast to $^{13}$CO$_2$, CO ice appears to be enhanced along the line of sight that traces envelope matter in front of the outflow cavities. We measure peak values of $N$ $\sim$ 4 $\times$ 10$^{18}$ cm$^{-2}$ and peak optical depths of $\tau$ $\sim$ 5. The column density decreases in the disk lane to $\sim$ 2.2 $\times$ 10$^{18}$ cm$^{-2}$ and the peak optical depths measured in this region are on average $\tau$ $\sim$ 2.

At first glance this points towards a decrease of CO ices in the central regions where the disk is located, however, a closer look at the band profiles between positions C, D, and E of Figure \ref{fig:L1527_int} shows that this is likely an effect of lower continuum in the dark disk lane. The continuum level decreases by almost an order of magnitude in this dark disk region, this can be seen in the upper right panel of Figure \ref{fig:L1527_int} where the spectra are shown on flux scale for the positions A, B, C and F. As a result of the lower continuum level, strong ice features such as CO are lowered into the noise floor. Consequently, the bottom parts of the band are not observed resulting in smaller integrated optical depths and thus lower column densities in this disk region compared to the regions towards outflow cavity where spectra have higher continuum levels. On average, the overall column densities agree with the CO ice column density of $N$ $\sim$ $\times$ 10$^{18}$ cm$^{-2}$ derived in \citet{brunken2024b} from aperture extractions at central position, but the weaker continuum in the disk region results in slightly lower column densities compared to other regions in the envelope which in turn is observed as the dark lane in Figure \ref{fig:L1527_CD} panel II.  %U NEED TO CORRECT BRUNKEN VALUES WITH THE NEW BANS STRENGHTS

\subsubsection{OCN$^-$}

In Figures \ref{fig:L1527_CD} and \ref{fig:L1527_OD} panels III we present the column density and peak optical depth maps of the 4.60 $\mu$m OCN$^-$ ice feature. Similar to $^{13}$CO$_2$, OCN$^-$ ice appears to be enhanced in the disk regions with column densities reaching $N$ $\sim$ 2 $\times$ 10$^{17}$ cm$^{-2}$ peak optical depths of $\tau$ $\sim$ 0.7. A growth of the ice feature is observed between positions B and C closer to the disk lane (Figure \ref{fig:L1527_int}) . 

\begin{figure*}[h!]
    % \vspace{0.5cm}
    \centering
     \includegraphics[width=0.8\hsize]{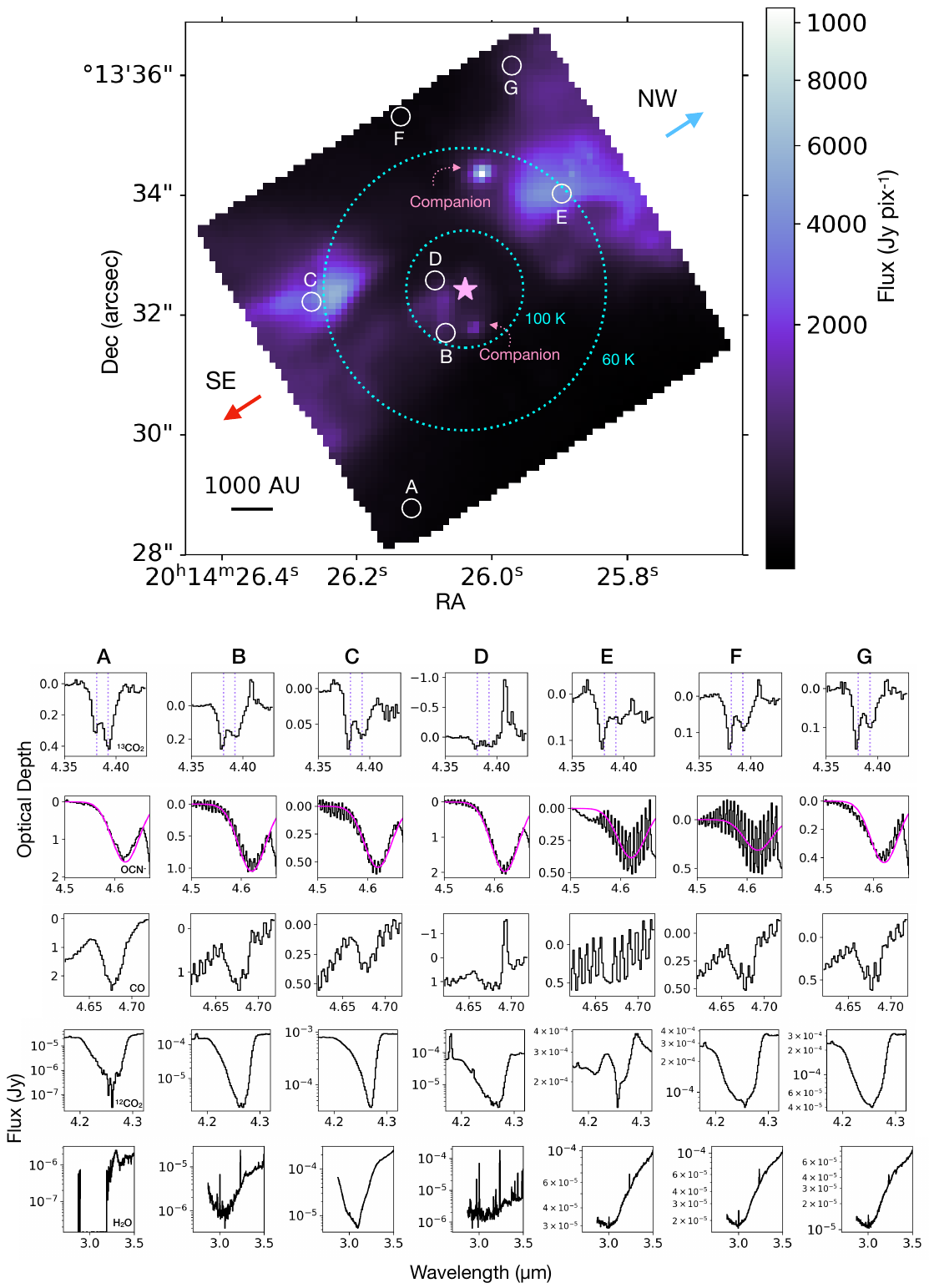}
    \caption{Intensity map of IRAS 20126 at 3.8 $\mu$m. The source location is marked with a star and the location of the companion is marked with pink arrows. The white circles are centered on the pixel where spectra were extracted with the corresponding spectra of $^{13}$CO$_2$, CO, OCN$^{-}$, $^{12}$CO$_2$ and H$_2$O shown in the panels below. The 60 K and 100 K dotted circles mark the temperature borders derived from a simple 1D spherical radiative transfer model. The dotted purple lines in the first row show the locations of the 4.38 $\mu$m and 4.39 $\mu$m of peaks of the $^{13}$CO$_2$ ice feature. The pink solid lines show the Gaussian curves fitted over the OCN$^-$ ice feature. The blue and red arrows show the directions of the outflow. }
    \label{fig:IRAS_20126_int}
\end{figure*}

Towards the outflow cavities, the ice column densities are on average lower with $N$ $\sim$ 9 $\times$ 10$^{16}$ cm$^{-2}$. We note that northern and southern regions of the map appear to show enhanced ice column densities but this is caused by artifacts in the spectra due to the low signal levels. We also note that a number of darker pixels are observed in the most central regions of the disk and that these are also the result of noise spikes in the spectra that complicates integration under the ice feature. The column densities on average are consistent with the column density of 8.1 $\times$ 10$^{16}$ cm$^{-2}$ derived by \citet{nazari2026} from aperture extractions on source position when using the same band strenghts.

\begin{figure*}[h!]
    % \vspace{0.5cm}
    \centering
     \includegraphics[width=0.7\hsize]{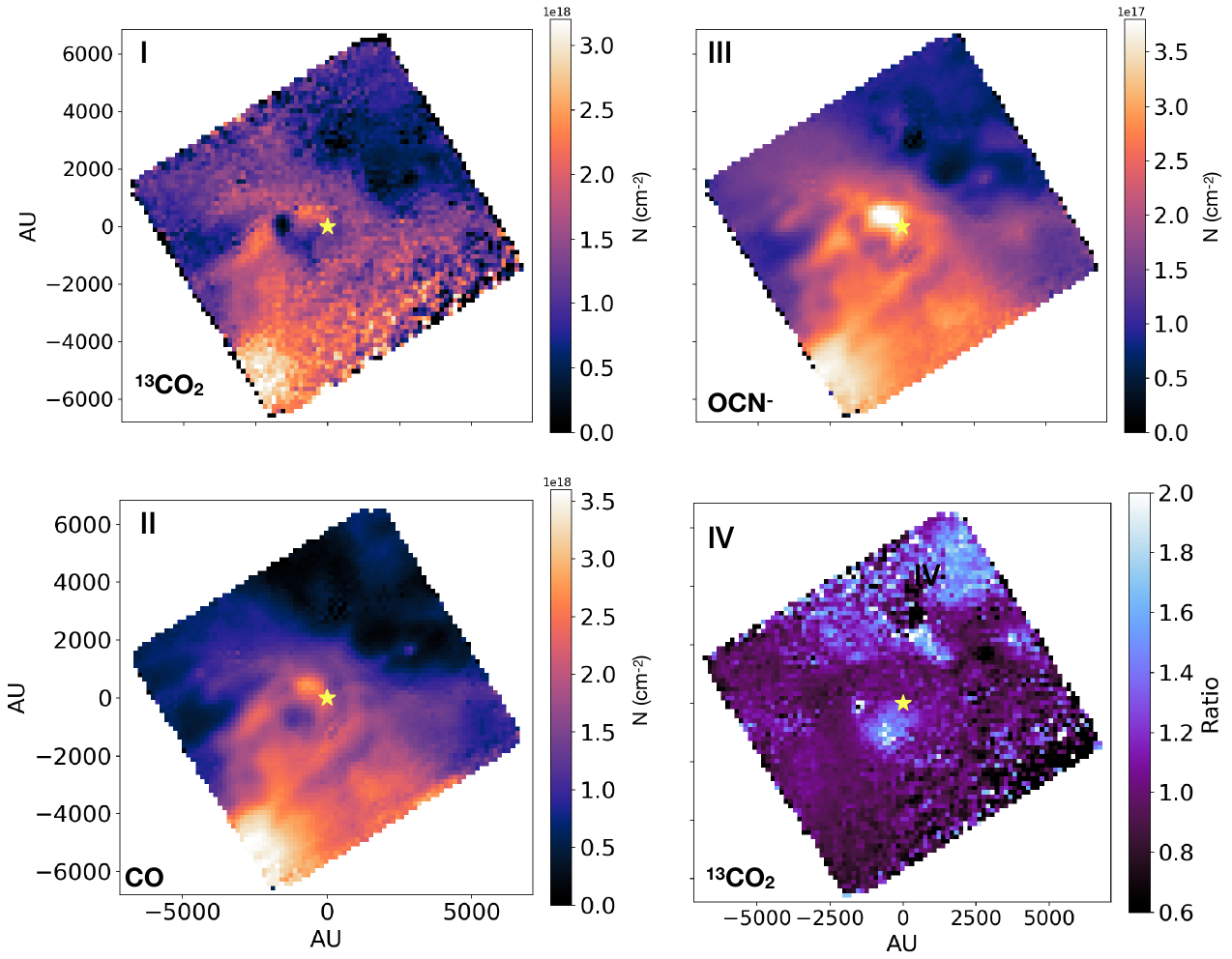}
    \caption{Column density maps of $^{13}$CO$_2$, CO and OCN$^{-}$ ice (panel I, II and III, respectively). $^{12}$CO$_2$ column densities are derived from $^{13}$CO$_2$ column densities. Panel IV shows the ratio between the 4.38 $\mu$m and 4.39 $\mu$m peak of the $^{13}$CO$_2$ ice feature. A higher ratio indicates higher degree of CO$_2$ ice segregation. The axes are shown on astronomical unit scales.}
    \label{fig:IRAS20126_CD}
\end{figure*}
\subsubsection{Ice column density ratios}

Ice column density ratio maps of CO$_2$/CO,  CO$_2$/OCN$^-$ and OCN$^-$/CO are presented in Figure \ref{fig:L1527_abun} panels I, II and III. The CO$_2$ ratio maps are derived from the $^{12}$CO$_2$ ice column densities discussed in the previous section. The abundance of CO$_2$ ice relative to CO increases considerably in the disk regions as shown in panel I. This is a result of both the lower column densities measured for CO ice, likely due to the bitting hitting the noise floor, and the enhancement of CO$_2$ ices in these central regions and should therefore be regarded as an upper limit. Panel II shows that CO$_2$ is also slightly enhanced in the disk region compared to OCN$^-$. Finally, panel III shows less CO ices relative to OCN$^-$ in the central regions, but this is again likely a lower limit due to the weaker continuum level.

\begin{figure*}[h!]
    % \vspace{0.5cm}
    \centering
     \includegraphics[width=1\hsize]{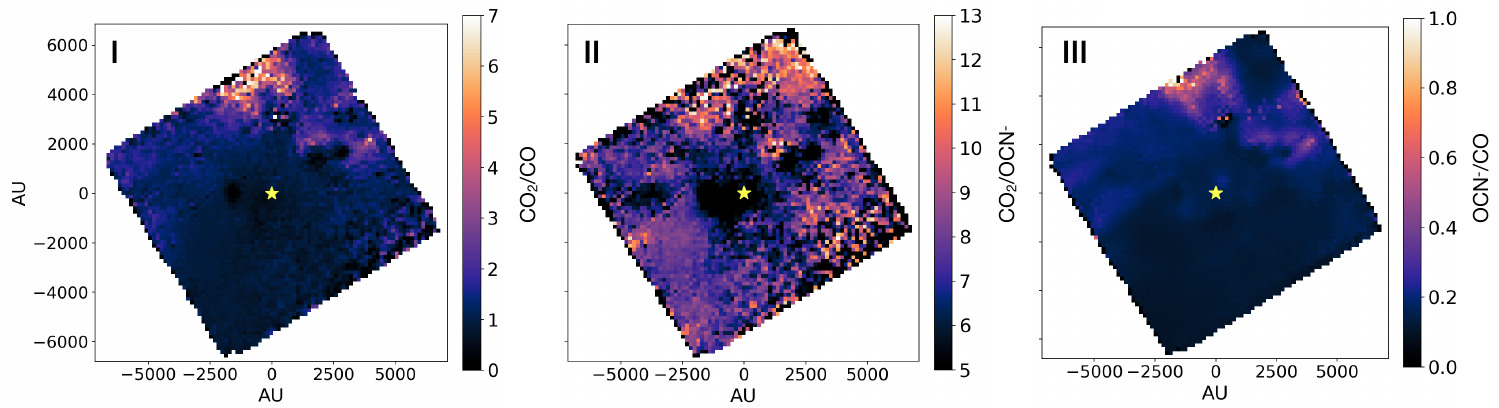}
    \caption{Column density ratio maps for CO$_2$, OCN$^-$ and CO ice towards IRAS 20126. The $^{12}$CO$_2$ column densities were determined using the standard $^{12}$C/$^{13}$C ratio of $\sim$ 68.}
    \label{fig:IRAS_abun}
\end{figure*}

\subsection{IRAS 20126}

In Figure \ref{fig:IRAS_20126_int} we present the 3.8 $\mu$m intensity map of IRAS 20126. The central source is embedded in a disk and two outflows stretch out in the NW and SE directions \citep{cesaroni2025,federman2023}. Spectra taken at various positions in the envelope are shown on optical depth scale in panels A to G for $^{13}$CO$_2$, OCN$^{-}$ and CO ice and on flux scale for $^{12}$CO$_2$ and H$_2$O. The full 2.9 - 5 $\mu$m spectra taken at the various positions are presented in Figure \ref{fig:IRAS20126_full}.

% \begin{figure}[h!]
%     % \vspace{0.5cm}
%     \centering
%      \includegraphics[width=1\hsize]{IRAS20126-intensity.pdf}
%     \caption{Intensity map of the 3.8 $\mu$m continuum of IRAS 20126 in mJy/beam. The centra star is marked with a pink star. A NW and SE outflows ar }
%     \label{fig:IRAS_20126_int}
% \end{figure}

\subsubsection{CO$_2$}

The column density map of $^{12}$CO$_2$ ice and the peak optical depth map of $^{13}$CO$_2$ ice are shown in panels I of Figures \ref{fig:IRAS20126_CD} and  \ref{fig:IRAS20126_OD}, respectively. Both maps show a high degree of non-uniformity in the ice distribution. The southern region of the envelope in particular appears to be enhanced in CO$_2$ ices with column densities reaching up to $N$ $\sim$  3 $\times$ 10$^{18}$ cm$^{-2}$ ($^{12}$CO$_2$) and peak optical depths of $\sim$ $\tau$ $\sim$ 0.4 ($^{13}$CO$_2$). The $^{13}$CO$_2$ spectrum taken at position A (Figure \ref{fig:IRAS_20126_int}) indeed shows a stronger ice feature compared to other positions in the envelope where the optical depths can range between $\tau$ $\sim$ 0.1 - 0.2 (positions C, F and G). 

In addition, we note that the long wavelength $^{13}$CO$_2$ peak (4.39 $\mu$m) is considerably stronger compared to the short wavelength peak (4.38 $\mu$m) at position A, implying that segregation might have just started. This trend reverses at other positions in the envelope, for example, at positions B, C, E, F, and G. Spectra taken at positions F and G in particular show a significantly enhanced 4.38 $\mu$m and this is also reflected in the ratio map in Figure \ref{fig:IRAS20126_CD} (panel IV). A high ratio is also observed in the central regions of the maps, slightly south of the location of the central source (position B).  

In the fourth row of Figure \ref{fig:IRAS_20126_int} we show the spectra of the 4.27 $\mu$m stretching mode of $^{12}$CO$_2$ on flux scale. The high S/N of the data allows us to observe this ice band in its entirety at a few positions, a feat that is not usually possible because this feature often saturates into the noise level due to its strength. Clear variations are seen in the band profile, the most notable being the additional sharp spectral feature that appears at positions E. This additional absorption peak is likely the ice absorption feature of segregated pure CO$_2$ ice , consistent with the profile of the $^{13}$CO$_2$ feature at this position. The continuum is also significantly warped by scattering effects at position E \citep{dartois2006spectroscopic,dartois2022influence, dartois2024}.

Finally, there are regions in the envelope where the CO$_2$ ice column density decreases considerably. These regions appear to coincide with regions where the ratio between the two $^{13}$CO$_2$ peaks is low, indicative of colder ices. Indeed, Panel I appears to be almost the opposite of panel IV (Figure \ref{fig:IRAS20126_CD}). Finally, we observe several "dark spots" in the map. These likely coincide with shock knots or the positions of the two stellar companions whose outflows could be interacting with the envelope material of IRAS 20126 \citep{federman2023}. On average, the CO$_2$ column densities are consistent with the value of 2.1 $\times$ 10$^{18}$ cm$^{-2}$ derived from aperture extractions on source position \citep{brunken2025}

\subsubsection{CO}

The column density and peak optical depth maps of CO ice are shown in Figures \ref{fig:IRAS20126_CD} and \ref{fig:IRAS20126_OD} in panels II, respectively. CO follows CO$_2$ closely with higher column densities ($N$ $\sim$ 3.5 $\times$ 10$^{18}$ cm$^{-2}$) measured in south regions of the envelope. The continuum of IRAS 20126 is strong and the CO band is therefore well above the noise floor, meaning that the bottom of the ice band is not hitting the noise floor, as is the case for L1527. Changes in the gaseous CO line forest are also observed at the different positions. The lines grow particularly strong at position E (Figure \ref{fig:IRAS_20126_int}), a signature of shocked gas and thus high temperatures. This is the same position where the CO ice band grows significantly weaker. Dark spots and patches are also observed in the CO maps, most of which coincide with the spots already seen in the CO$_2$ maps.

\subsubsection{OCN$^-$}

In Figures \ref{fig:IRAS20126_CD} and \ref{fig:IRAS20126_OD} panels III we present the column density and peak optical depth maps of OCN$^-$. The most notable difference between the OCN$^-$ map and those of $^{13}$CO$_2$ and CO is the bright region near the central source position. Figure \ref{fig:IRAS20126_CD} shows that the band indeed grows ($\tau$ > 1) at position D. The peak column densities reach values of $\sim$ 4 $\times$ 10$^{17}$ cm$^{-2}$ which are in agreement with the values \citet{nazari2024} derived for several warm Class 0/I sources.

\subsubsection{H$_2$O}

In the final row of Figure \ref{fig:IRAS_20126_int} we present the spectra of the 3 $\mu$m H$_2$O ice feature on flux scale taken at a number of positions in the envelope. Because a portion of the band is missing, we are unable to fit a continuum and provide a quantitative analysis. 

Even so the spectra show clear changes in the H$_2$O band profile. The known peaked crystalline feature of H$_2$O in particular is clearly observable at position C. This profile is not seen at positions E, G and F but this could be due to contamination from foreground emission that flattens the bottom of the feature \citep{slavicinska2024}. At positions A and D, the feature saturates into the noise level, and therefore a proper diagnostic of the band shape is not possible at every location in the envelope.

\subsubsection{Ice column density ratios}

The ice column density ratio maps of CO$_2$/CO,  CO$_2$/OCN$^-$ and OCN$^-$/CO ice are presented in panels I, II and III of Figure \ref{fig:IRAS20126_CD}. The abundance of both CO$_2$ and OCN$^-$ appear to increase relative to CO in the NE region of the envelope, coinciding with the locations where the CO ice feature decreases considerably in strength (Figure \ref{fig:IRAS_20126_int}). Enhancement of CO$_2$ relative to OCN$^-$ is observed mostly in the NW region where the $^{13}$CO$_2$ feature is segregated. In contrast, OCN$-$ appears to be more enhanced near the central source also where high CO$_2$ segregation is also observed as shown in panel N of Figure \ref{fig:IRAS20126_CD}.

% The column density ratio between CO$_2$ and OCN$^-$ ice also increases in the upper regions of the envelope where the strength of the OCN$^-$

\section{Discussion}
\label{sec:4}

The results in Section \ref{sec:3} indicate that there is a significant variation in the distribution of CO$_2$, CO,  H$_2$O and OCN$^-$ ices in the envelopes of both L1527 and IRAS20126. The temperature structure of these two envelopes is well documented in the literature, which is imperative for interpreting these findings, since ices are highly susceptible to their local environment. 

Ices form during the cloud stage through the adhesion of atoms and small molecules to the surfaces of cold dust grains, after which they evolve in tandem with the protostellar system. Hydrogen-rich molecules such as H$_2$O, NH$_3$ and CH$_4$ are the first to form due to the high abundance of hydrogen atoms during the translucent cloud stage \citep{oberg2011spitzer,boogert2015observations}. CO$_2$ is also produced in this H$_2$O-dominated ice via the HO-CO complex \citep{oba2010experimental, ioppolo2011surface, garrod2011}. This epoch is closely followed by the heavy freeze-out stage, where CO-dominated ices grow rapidly to form an apolar ice mantle \citep{pontoppidan2006}. A fraction of this CO ice is converted into CO$_2$, and hydrogenation of CO also leads to the formation of larger molecules such as CH$_3$OH \citep{cuppen2024}. The result is a stratified ice structure, where the molecules are intimately mixed in the different ice layers.

Naturally, when these ices are processed by either thermal or UV radiation, the different species will react according to their intrinsic properties or undergo further chemical reactions to form other molecules. The highly volatile CO ice, for instance, will partially desorb from the apolar ice mantle. Further heating will cause restructuring of the molecular bonds and segregation of molecules such as CO$_2$, which produces inclusions of pure CO$_2$ ice within the mantle, as well as crystallization of the ices. 

These structural changes are then reflected in the infrared ice absorption bands with CO$_2$ showing a double peak profile at 2.70 $\mu$m , 4.39 $\mu$m and 15.2 $\mu$m \citep{boogert1999iso,pontoppidan2008c2d,poteet2013anomalous,brunken2024a,brunken2025} and crystalline water showing a characteristic peaked profile at 3 $\mu$m \citep{fraser2001}. These unique spectral features are signatures of ice processing and can offer vital clues about the local environment. 
In the following sections, we examine the structural difference between L1527 and IRAS 20126 and explore a possible link between the architecture of the two systems and the ongoing ice chemistry. 

\begin{figure}[h!]
    % \vspace{0.5cm}
    \centering
     \includegraphics[width=1\hsize]{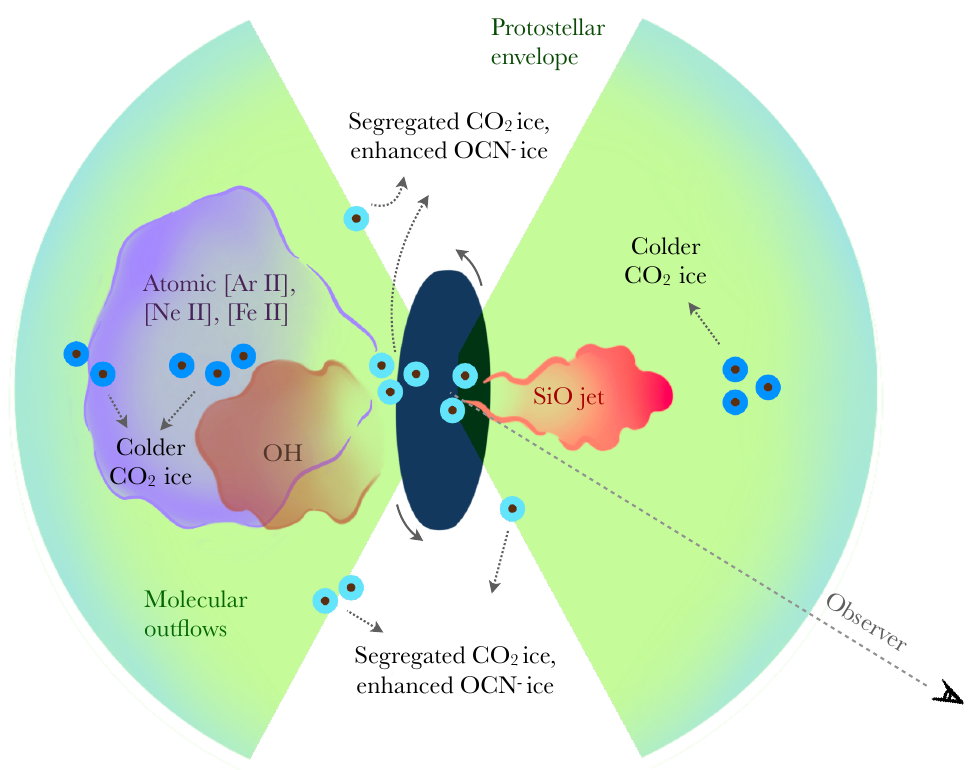}
    \caption{Schematic of the L1527 system. The molecular outflows observed with JWST NIRSpec. The SiO jets in L1527 \citep{hoff2023} are illustrated in red while the disk is depicted in blue. The atomic and ionic emissions detected by \citet{drechsler2026} in L1527 are depicted in purple and brown, respectively. The icy grains provide a summary of the main findings in this work.}
    \label{fig:L1527-schematic}
\end{figure}

\subsection{Ice distribution in L1527}

L1527 is a young Class 0 protostar located at a distance of about 142 pc \citep{torres2007,tobin2012}. The central protostar has a luminosity between 2 - 3  L$_{\odot}$ \citep{tobin2008,kristensen2012} and is enshrouded in a near edge-on disk  ($i$ = 85$^{\circ}$) that is undergoing Keplerian motion \citep{tobin2008, ohashi2014, hoff2023}. The envelope contains about 1 M$_{\odot}$ of material, a fraction of which is currently accreting onto the protostar. Finally, a bipolar outflow cavities are observed in the form of reflection nebulae on the east and west of the central source, which are carving out large cavities in the envelope material \citep{tobin2008, tobin2012}.

Due to its close proximity, L1527 offers an ideal setting for investigating the physical structure of young low mass protostars. A study by \citet{tobin2012}, showed clear asymmetries and variability in the outflow cavities where one lobe appears to be brighter than the other, and the authors attributed this to a probable misalignment of the central disk. More recent studies by \citet{drechsler2026} and \citet{devaraj2026} revealed asymmetric atomic emission coming from the outflow, with an enhancement observed in the east lobe, which points to an asymmetric mass accretion process. This asymmetry is also observed in Figure \ref{fig:L1527_int} where the eastern outflow appears brighter in the intensity map. \citet{devaraj2026} also reported the presence of a high velocity ionized jet coming from  Detailed modeling of the temperature structure was also conducted by \citet{hoff2018} and \citet{hoff2023} who derived $T_{\rm{disk}}$ $\sim$ 20 - 40 K. The authors showed that $^{13}$CO and C$^{18}$O gas emission are detected throughout the disk, while N$_2$D$^+$, which is destroyed by CO, is not detected, thus concluding that the disk is warm enough to prevent CO from freezing out. In addition to various other molecular tracers, they also detect SiO emission west of the central source, alluding to the presence of a jet. In Figure \ref{fig:L1527-schematic} we present an illustrated schematic of the envelope around L1527. The main gas tracers and the main ice findings of the study are highlighted at different locations in the envelope.

The ice maps in Figure \ref{fig:L1527_CD} imply a higher abundance of ices in the disk region and overall less ices towards the outflow cavities, as expected. More importantly, the appearance of the pure CO$_2$ ice feature at 4.38 $\mu$m shows that CO$_2$ is segregating in the regions close to the disk lane. The peak grows towards the center of the envelope and appears to trace the cavity walls on both sides of the disk (Figure \ref{fig:L1527_CD} panel IV). This indicates that these regions have high temperatures that prompt ice heating, and is consistent with the location where \citet{hoff2023} detects warm ($\geq$ 60 K) H$_2$CO 3$_{2,2}$ - 2$_{2,1}$ emission. Although the gas and dust do not necessarily have to be coupled, in these dense regions it can be assumed that the warm temperatures measured in the gas phase translate into a warm region overall that heats the ices. In fact, Figure \ref{fig:L1527_int} shows that the location where the pure CO$_2$ peak becomes prominent falls around the 60 K snowline, the typical temperature ranges where CO$_2$ would normally segregate \citep{ehrenfreund1998ice}. 
 
The OCN$^-$ ice column densities also appear to be enhanced in the disk lane (Figure \ref{fig:L1527_CD} panel III). Figure \ref{fig:L1527_CD} panel IV shows high segregation of CO$_2$ ice at these locations, indicating that observed enhancement of OCN$^-$ could be due to elevated temperatures. This is consistent with the study by \citet{nazari2026} who showed that the OCN$^-$ ice feature can be enhanced up to a factor 2 - 3 in warmer environments (see also \citet{broekhuizen2005}). 
Finally, Figure \ref{fig:L1527_CD} panel II indicate an opposite trend for CO where the CO ice appears to be largely present in the cavity walls but is deficient in the disk. Although the warm disk scenario presented by \citep{hoff2023} could explain the lack of CO ices in the disk lane, we must also consider the weak continuum in these regions, as mentioned in Section \ref{sec:3} that would cause the CO ice band to hit the noise level (Figure \ref{fig:L1527_int}). Therefore, while we do not completely discard the possibility that the lower CO ice column densities in the disk lane are due to ice desorption, we also cannot say conclusively that this is the only reason given the weak signal in this region.

\begin{figure}[h!]
    % \vspace{0.5cm}
    \centering
     \includegraphics[width=1\hsize]{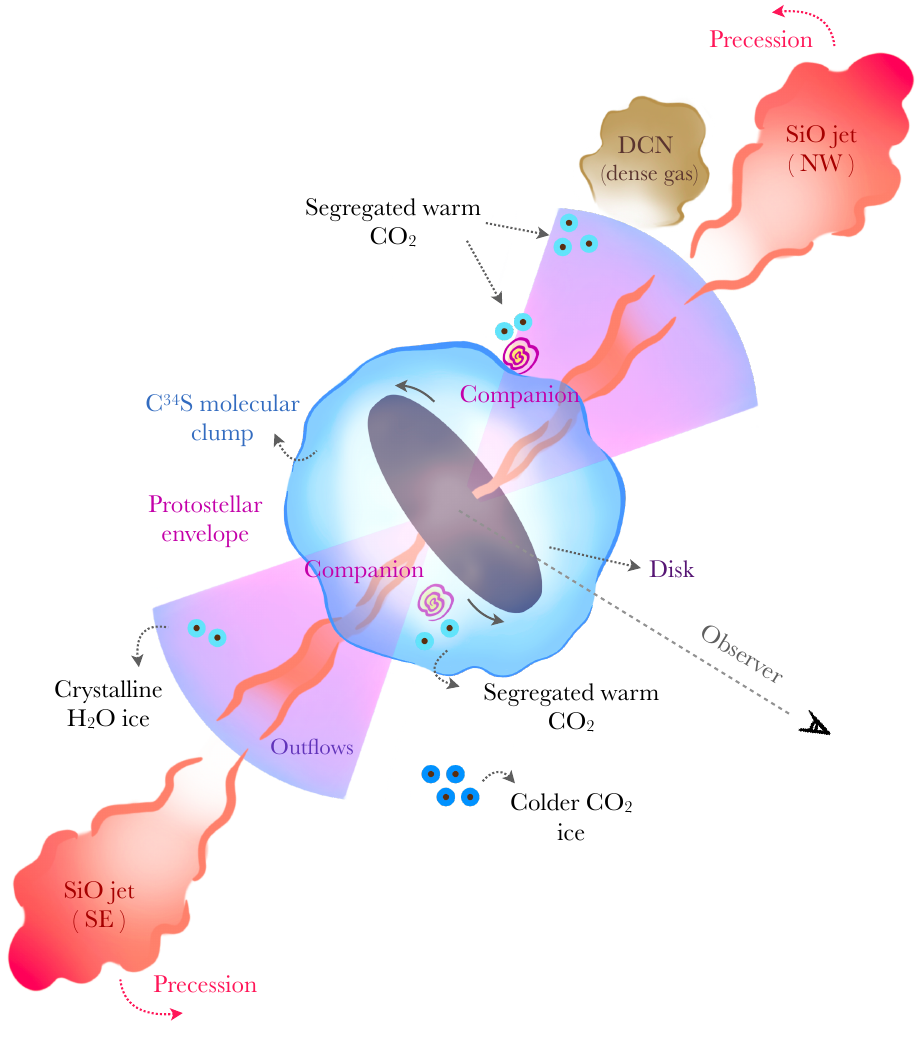}
    \caption{Schematic of the IRAS 20126 system. The molecular outflows observed with JWST NIRSpec. The SiO jets in IRAS 20126 \citep{cesaroni2025} are illustrated in red while the disk with the molecular C$^{34}$S clump \citep{cesaroni1997} is depicted in blue. The dense gas in the NW region traced by DCN \citep{cesaroni2025} is shown in brown. The positions of the young stellar companions in IRAS 20126 are illustrated with two knots in pink. Finally, the icy grains provide a summary of the main findings in this work.}
    \label{fig:IRAS20126-schematic}
\end{figure}

\subsection{Ice distribution in IRAS 20126}

IRAS 20126 is a known Class 0 high mass protostar that has been studied in both gas and ice phase \citep{cesaroni1997,cesaroni1999, federman2023, brunken2024a, slavicinska2024, brunken2025}. The system is located at 1.6 kpc \citep{moscadelli2011} and has a luminosity of 10$^4$ L$_{\odot}$ \citep{cesaroni2023}. The central star is embedded in a Keplerian disk \citep{cesaroni2005} that together with the protostar drive powerful molecular outflow. Two jets have been observed in SiO, H$_2$ and CO emission expanding in the NW and SE directions and their morphology suggests that they could be undergoing precession \citep{shepherd2000,cesaroni2005,lebron2006,caratti2008}. [Fe II] emission revealed the presence of compact knots and substructures in the outflows \citep{federman2023} and two companions are observed north and south of the outflow \citep[and references therein]{federman2023}. Finally, \citet{cesaroni1997} used C$^{34}$S to show that the disk is embedded in a 40 K molecular clump. Figure \ref{fig:IRAS20126-schematic} shows a schematic of the morphology of the IRAS 20126 system, including both gas-phase results and the main findings of the ice observations. 

Warm ices are detected throughout the entire envelope of IRAS 20126. This can be concluded from the presence of the 4.38 $\mu$m pure CO$_2$ peak observed at all locations in the envelope (Figure \ref{fig:IRAS_20126_int}). However, it is possible to distinguish between locations of higher temperature. Figure \ref{fig:IRAS20126_CD} panel I for example shows that the ices in the southern region of the envelope appear to be enhanced and that the ratio between the short wavelength and long wavelength peak of the $^{13}$CO$_2$ ice feature is lower in this region compared to every other location in the envelope (Panel N). There is also a lack of gaseous CO forest line at this location. This suggests that the ices are generally less heated at this location. A dense region of high extinction where the material is generally cooler could be a possible explanation for the colder ices. Another possible explanation is disk-shadowing, where the disk shields the central protostar, casting shadows on the envelope, ultimately resulting in regions with lower temperatures \citep{murillo2015}. However, to confirm this scenario, observations of deuterated species such as DCO$^+$ are needed, since their production is enhanced under these cooler conditions.

The possibility that these colder ices could be originating from the 40 K molecular clump that \citet{cesaroni1997} observed in C$^{34}$S emission was also considered. In the laboratory CO$_2$ segregation usually occurs between 50 - 90 K \citep{ehrenfreund1998ice,dartois1999} but under astrophysical conditions the longer time scales mean that thermal effects can take place at lower temperatures \citep{boogert2008}. At 40 K the ices could therefore be warm enough to begin the process of distillation and segregation, resulting in the double peak $^{13}$CO$_2$ profile, while CO remains partially trapped in the ice mantle along with the less volatile species. As shown in Figure \ref{fig:IRAS20126_overlays} (right panel) however, the region where we detect the colder ices does not coincide with the location of the molecular clump.

The appearance of strong pure CO$_2$ peaks at 4.38 $\mu$m and 4.27 $\mu$m in the NW outflow (at for example positions E and F), together with the strong gaseous CO line forest indicate significant ice heating and ice segregation at these locations (Figure \ref{fig:IRAS_20126_int}). We also note the low abundance of CO ice at these locations (Figure \ref{fig:IRAS20126_CD} panel I) and the enhancement of both CO$_2$ and OCN$^-$ relative to CO in this region (Figure \ref{fig:IRAS_abun} ) that suggest sublimation of CO ice due to higher temperatures. These elevated temperatures are likely the result of interaction between dense envelope material and the NW jet, which in turn warms the environment and heats the surrounding material. Indeed, \citet{cesaroni2025} traced the jets in both SiO and H$_2$CO and showed that the SE H$_2$CO jet appears to be cooler and less dense, while the NW jet is still passing through denser material. In Figure \ref{fig:IRAS20126_overlays} (right panel) we show an RGB JWST NIRSpec and ALMA composite image. The DCN map \citep{cesaroni2025} traces regions of dense material and the contours show the locations of the NW and SE SiO jets \citep{cesaroni2025}. It is evident from the DCN map that the NW part of the envelope contains dense material close to where the molecular outflow is observed with NIRSpec and where the SiO jet is located. Furthermore, \citet{shepherd2000,cesaroni2005,lebron2006,caratti2008} demonstrate that the jet is precessing and curving upward in the NW direction. Therefore, we can conclude that the NW region of the envelope is undergoing significant disturbance, likely creating a localized region of elevated temperature and possibly radiation that could be heating the ices.

Another region of high temperatures is seen close to the central source as shown in Figure \ref{fig:IRAS20126_CD} panel IV. The $^{13}$CO$_2$ is also not only segregated but we also observe enhancement of the OCN$^-$ relative to CO$_2$ (Figure \ref{fig:IRAS_abun} panel II) which again agrees with the findings in \citet{nazari2026}. Why this same enhancement is not observed in the NW warm region is however unclear. Finally, we detect crystalline water in the east envelope which indicates high temperatures at these locations (> 100 K). \citet{tyagi2025} also observed crystalline water in the intermediate mass source HOPS 370 in the vicinity of a shock knot, showing that these high temperatures are not uncommon in these massive sources. 

\subsection{High-mass vs. Low-mass}
From the previous sections, we note a few notable differences between the envelopes of the low-mass protostar L1527 and the high-mass protostar IRAS 20126. The geometry of L1527 is simpler compared to IRAS 20126 and the heating of ices appears to occur in a more structured manner compared to the high-mass counterpart. Ice segregation occurs gradually as a fuction of decreasing radius with a high temperature zone around the disk regions as shown with the thermal tracer $^{13}$CO$_2$ and the slightly enhanced OCN$^-$. IRAS 20126, on the contrary,  has a complex system that consists of outflow and jets that interact with the envelope medium. Heated ices are detected in front of the outflow cavities with particularly high temperature zones observed where the jet and outflow interact with envelope matter. 

One possible explanation for the heated conditions in the envelope of IRAS 20126 could also be its variability. Indeed, previous studies by \citet{massi2023,szymczak2024} and \citet{cesaroni2026} have shown that IRAS 20126 is a variable source, likely due to episodic accretion bursts. These periodic luminosity outbursts can flash-heat the ices on very short timescales, contributing to the elevated temperature zones observed in the ice maps. Furthermore, source geometry and substructures in the envelope can further cause non-uniform heating during outburst which could explain the non-uniform distribution of ices that are seen. Substructures in the envelope can also result in regions of high extinction where colder ices can survive, as seen in the southern regions of the envelope. 

This non-uniformity observed in IRAS 20126 could also be tied to the natal environments of the massive star, which can differ from that of low-mass systems. \citet{beuther2025} for instance argues that radiation and ionization pressure, occurring almost uniquely in high-mass star-forming regions, can impact the environment of the star. Moreover, as noted before, the presence of companions (Figure \ref{fig:IRAS_20126_int}) that drive additional outflows could be interacting with the envelope material of IRAS 20126 and altering its morphology. Finally, we also consider the age of the two systems where the slightly younger IRAS 202126 \citep{caratti2008} could still have remnants of the pre-stellar core that responsible for the substructure and non-uniform heating that are being observed.

Figures \ref{fig:L1527-schematic} and \ref{fig:IRAS20126-schematic} therefore offer a simplified interpretation of the two systems, but more studies are needed to further understand the architecture of these envelopes, with ices providing a complementary view of the gas and dust studies. This is especially the case for high-mass sources.

% well studied high-mass system protostar studi located 1.6 kpc. at  is a high-mass class 0 source located at   has been studied in detail across multiple wavelengths in both the gas and ice phase \citep{cesaroni1997,federman2023,brunken2025,cesaroni2025} and references therein. It is a high-mass accreting protostar that is surrounded by a disk undergoing keplerian rotation. \citet{cesaroni2025} traced the disk with CH$_3$CN and DCN and derived temperatures of $\sim$ 200 - 400 K. Jets traced by SiO and H$_2$CO are observed in the NW ans SE directions and molecular clump with T $\sim$ 40 K is detected in C$^{34}$S emission \citep{cesaroni1997,cesaroni2025}. A high-density region is also observed in the NW part in DCN emission as well as H$_2$CO \citep{cesaroni2025}. The NW and SE outflows are also observed in scattered light with JWST NIRSpec IFU where two companions are observed in Br$\alpha$ emission and shock knots are observed in Fe[II] \citep{federman2023}. Crystalline HDO ice \citep{slavicinska2024} and segregated CO$_2$ ice \citep{brunken2024a,brunken2025} are detected in the envelope signaling elevated temperatures and protostellar heating. In the following sections we present the spatial analysis of the ices the JWST NIRSpec IFU. 

\section{Conclusions}
\label{sec:5}

In this work we present JWST NIRSpec IFU ice spatial mapping in the envelopes of the low mass protostar L1527 and the high mass protstar IRAS 20126. This is the first time spatial mapping has been performed for a high mass source. Our main findings suggest that the distribution of ices in protostellar envelopes is directly tied to the temperature profile of the system and that substructures can exist for both low mass and high protostars.
\begin{itemize}
    \item The ices in L1527 are enhanced in the central regions where the disk is located, with lower column densities being measured towards the line of sight that traces envelope material in front of the outflow cavities where material is being carved out. 
    
    \item Segregated CO$_2$ warm ice is observed in L1527 in the vicinity of the disk and appears to trace the cavity walls. This points towards a warm region that also coincides with the warm disk layers observed in gas-phase studies. Enhancement of OCN$^-$ relative to both CO and CO$_2$ is observed at these warmer locations which is consistent with previous studies that showed that OCN$^-$ can be enhanced up to a factor 2-3 after its formation.

    \item The distribution of ices in the high mass protostar IRAS 20126 is less structured compared to L1527. The warm segregates and crystalline ices appear to be co-spatial with regions where material is being shocked in the NW outflow. In addition, segregated ices are seen in the warm region close to the central as source as well as enhancement of OCN$^-$ relative to the CO$_2$ due to the elevated temperatures. This enhancement of OCN$^-$ is however not observed in the NW shocked region.
    
    \item A region of relatively colder ices is detected in the southern region of the envelope of IRAS 20126. One possible explanation is disk shadowing but observations of deuterated species are needed to confirm this scenario. 

    \item Notable differences are observed between the envelope structure of the low-mass protostar L1527 and the high-mass protostar IRAS 20126. The there is a gradual heating of the ices in L1527 as a function of distance to the centra source with, higher segregation observed in the central disk region. The geometry and structure of IRAS 20126 on the other hand is far more intricate with warm ices detected in regions where outflow and jets interact with envelope medium, and as a result heat the surroundings. Dense regions of high extinction could be introducing additional substructres in the envelope of IRAS 20126 where ices are better shielded and resulting in colder ices.
    
\end{itemize}

Our study shows that the physical structure of protostellar envelopes is intimately linked with its ice chemistry. The distribution of ices in particular appear to be co-spatial with the temperature structure of the envelopes. The envelopes around high-mass protostar are far more complex than their low-mass counter parts and further studies focused on mapping are needed to better understand these systems and to search for similarities and trends.

\begin{acknowledgements}

Astrochemistry in Leiden is supported by the Netherlands Research School for Astronomy (NOVA), by funding from the European Re- search Council (ERC) under the European Union’s Horizon 2020 research and innovation programme (grant agreement No. 101019751 MOLDISK), and by the Dutch Research Council (NWO) grant 618.000.001. Support by the Danish National Research Foundation through the Center of Excellence “InterCat” (Grant agreement no.: DNRF150) is also acknowledged.

This work is based on observations made with the NASA/ESA/CSA James Webb Space Telescope. The data were obtained from the Mikulski Archive for Space Telescopes at the Space Telescope Science Institute, which is operated by the Association of Universities for Research in Astronomy, Inc., under NASA contract NAS 5-03127 for JWST.

The following National and International Funding Agencies funded and supported the MIRI development: NASA; ESA; Belgian Science Policy Office (BELSPO); Centre Nationale d’Etudes Spatiales (CNES); Danish National Space Centre; Deutsches Zentrum fur Luft- und Raumfahrt (DLR); Enterprise Ireland; Minis- terio De Economiá y Competividad; Netherlands Research School for Astron- omy (NOVA); Netherlands Organisation for Scientific Research (NWO); Sci- ence and Technology Facilities Council; Swiss Space Office; Swedish National Space Agency; and UK Space Agency.

NB thanks R. Cesaroni for providing additional data to support the observations. A.C.G. acknowledges support from PRIN-MUR 2022 20228JPA3A “The path to star and planet formation in the JWST era (PATH)” funded by NextGeneration EU and by INAF-GoG 2022 “NIR-dark Accretion Outbursts in Massive Young stellar objects (NAOMY)”, and by Large Gran INAF-2024 “Spectral Key fea-tures of Young stellar objects: Wind-Accretion LinKs Explored in the infraRed (SKYWALKER). PN is supported by a Warwick Prize Fellowship.”

\end{acknowledgements}

\bibliographystyle{aa}
\bibliography{main}

%\twocolumn

%\onecolumn

\begin{appendix}

\section{Supplementary figures}
\label{Appendix:A}

\begin{figure}[h!]
    % \vspace{0.5cm}
    \centering
     \includegraphics[width=1\hsize]{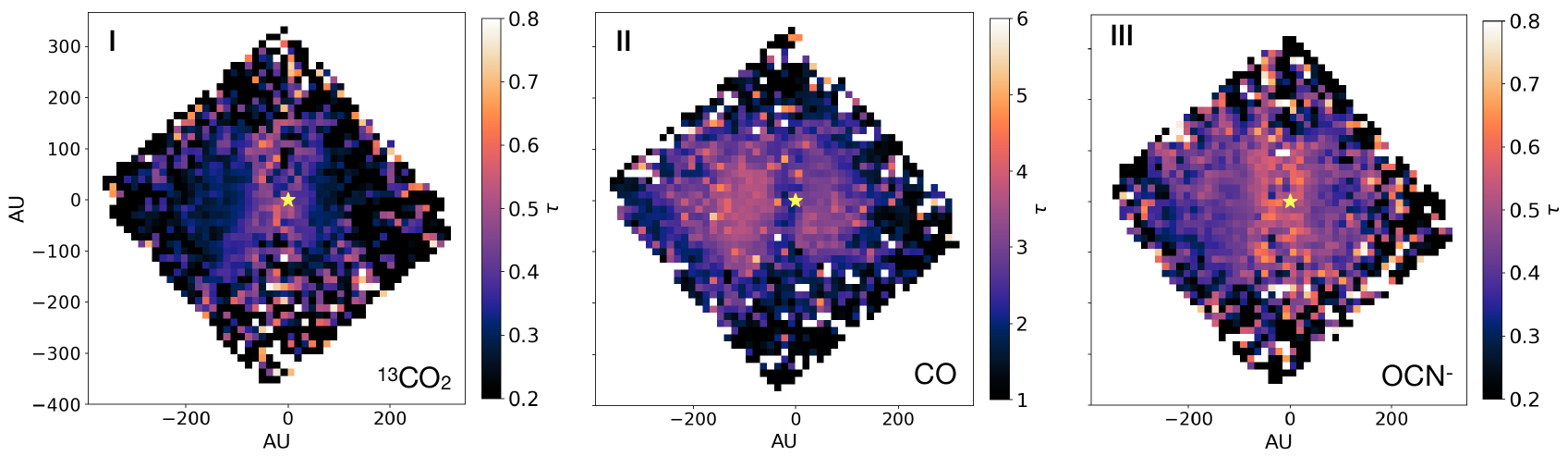}
    \caption{Peak optical maps of $^{13}$CO$_2$, CO and OCN$^{-}$ (panel A, B and C, respectively). }
    \label{fig:L1527_OD}
\end{figure}

\begin{figure}[h!]
    % \vspace{0.5cm}
    \centering
     \includegraphics[width=1\hsize]{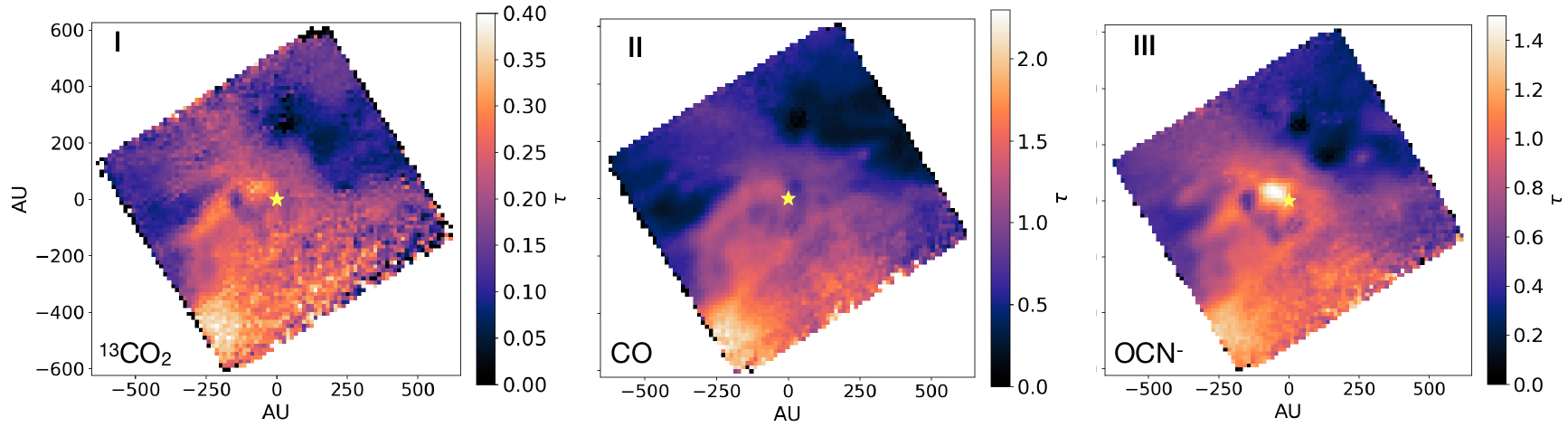}
    \caption{Peak optical maps of $^{13}$CO$_2$, CO and OCN$^{-}$ (panel A, B and C, respectively).}
    \label{fig:IRAS20126_OD}
\end{figure}

\begin{figure}[hbt]
    % \vspace{0.5cm}
    \centering
     \includegraphics[width=1\hsize]{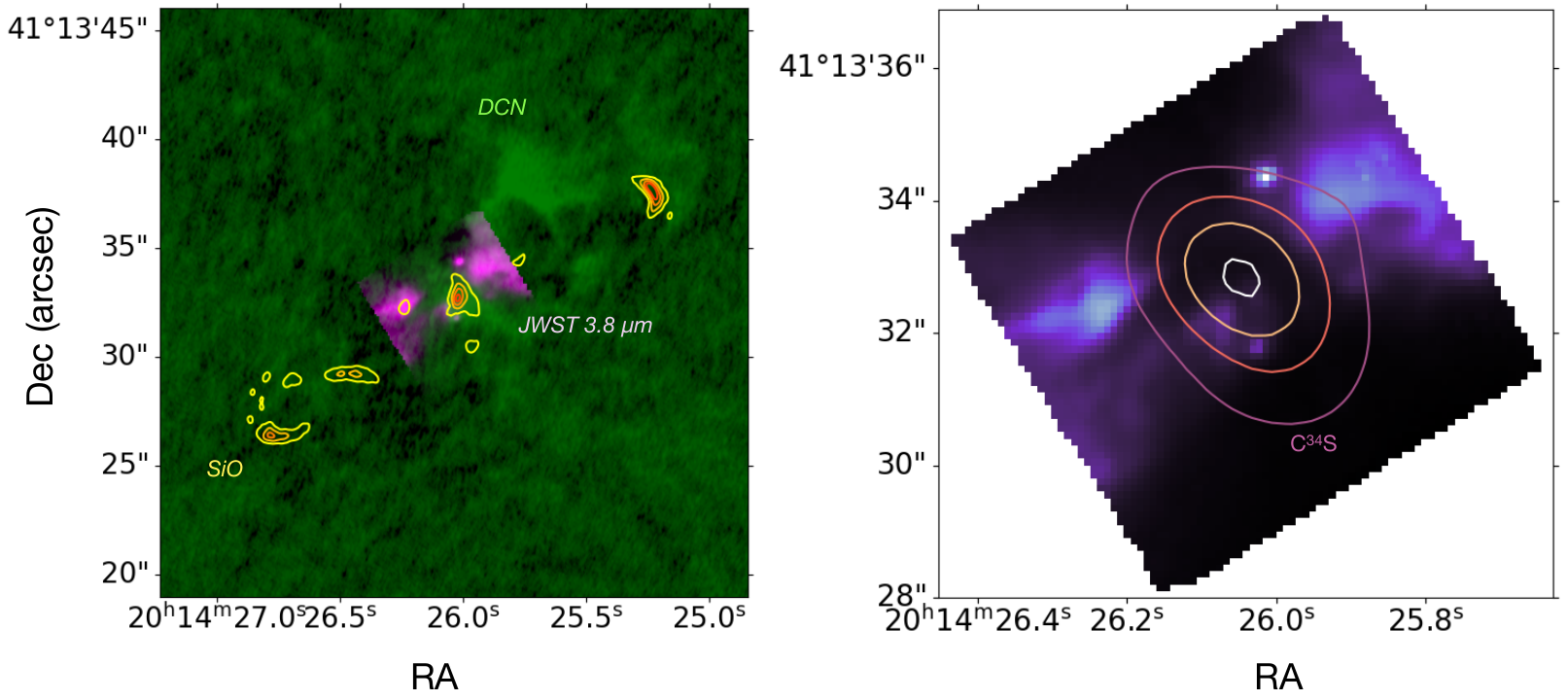}
    \caption{JWST ALMA composite images of IRAS20126. Panel A shows the NIRSpec 3.8 $\mu$m continuum (pink) with the SiO jet plotted as contours (yellow) and the DCN emission overlaid on top (green) \citep{cesaroni2025} Panel B shows the NIRspec 3.8 $\mu$m continuum with the CS emission overlaid as contours to show the location of the 40 K molecular clump discussed in \citet{cesaroni1997}.}
    \label{fig:IRAS20126_overlays}
\end{figure}

\begin{figure}[hbt]
    % \vspace{0.5cm}
    \centering
     \includegraphics[width=1\hsize]{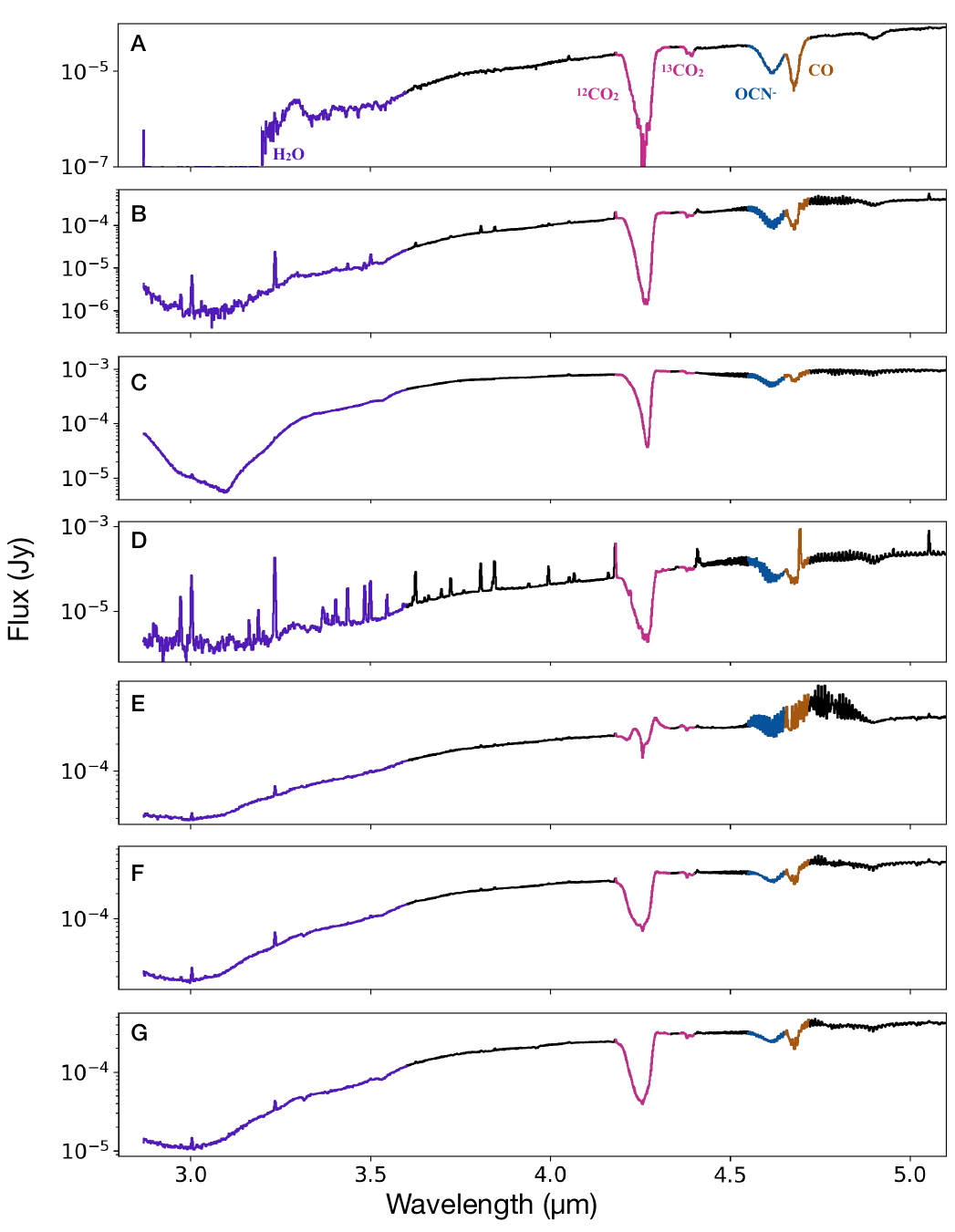}
    \caption{Full 2.9 - 5.3 $\mu$m spectra take at positions A to G in IRAS 20126. The ice bands are highlighted in color.}
    \label{fig:IRAS20126_full}
\end{figure}

\begin{center}
\begin{table}[hbt!]
\caption{Source parameters used in RADMC 1D spherical envelope models}
\small
\centering
% \begin{threeparttable}
\begin{tabular}{lcccccl}
\hline \hline
Source & R$_{env}$ (AU) & M$_{env}$ (M$_{\odot}$) & M$_{*}$ (M$_{\odot}$) & L$_{*}$ (L$_{\odot}$) & Distance (pc) \\ 
         
\hline

L157 & 10313 & 1 & 0.45  & 3 & 142  \\

IRAS 20126 & 115508 & 245  & 12 & 10$^4$ & 1550 \\

\hline
\end{tabular}
\label{Tab:1}
\begin{tablenotes}\footnotesize
\item{\textbf{Notes.} Source properties were taken from \citep{tobin2012, hoff2018, cesaroni2023}} 
\end{tablenotes}
% \end{threeparttable}
\end{table}  
\end{center}

\end{appendix}

\end{document}